# Tomogram-based Comparison of Geostatistical Models: Application to the Macrodispersion Experiment (MADE) Site

Niklas Linde[1]*, Tobias Lochbühler[1], Mine Dogan[2], Remke L. Van Dam[3,4]

[1]Applied and Environmental Geophysics Group, Institute of Earth Sciences, University of Lausanne, Lausanne, Switzerland;

[2]Department of Environmental Engineering and Earth Science, Clemson University, Clemson, South Carolina, USA;

[3]Science and Engineering Faculty, Queensland University of Technology, Brisbane, Australia;

[4]Department of Geological Sciences, Michigan State University, East Lansing, Michigan, USA.

*corresponding author: Niklas.linde@unil.ch

**Abstract**

We propose a new framework to compare alternative geostatistical descriptions of a given site. Multiple realizations of each of the considered geostatistical models and their corresponding tomograms (based on inversion of noise-contaminated simulated data) are used as a multivariate training image. The training image is scanned with a direct sampling algorithm to obtain conditional realizations of hydraulic conductivity that are not only in agreement with the geostatistical model, but also honor the spatially varying resolution of the site-specific tomogram. Model comparison is based on the quality of the simulated geophysical data from the ensemble of conditional realizations. The tomogram in this study is obtained by inversion of cross-hole ground-penetrating radar (GPR) first-arrival travel time data acquired at the MAcro-Dispersion Experiment (MADE) site in Mississippi (USA). Various heterogeneity descriptions ranging from multi-Gaussian fields to fields with complex multiple-point statistics inferred from outcrops are considered. Under the assumption that the relationship between porosity and hydraulic conductivity inferred from local measurements is valid, we find that conditioned multi-Gaussian realizations and derivatives thereof can explain the crosshole geophysical data. A training image based on an aquifer analog from Germany was found to be in better agreement with the geophysical data than the one based on the local outcrop, which appears to under-represent high hydraulic conductivity zones. These findings are only based on the information content in a single resolution-limited tomogram and extending the analysis to tracer or higher resolution surface GPR data might lead to different conclusions (e.g., that discrete facies boundaries are necessary). Our framework makes it possible to identify inadequate geostatistical models and petrophysical relationships, effectively narrowing the space of possible heterogeneity representations.

# 1. Introduction

Tomographic (geophysical and hydrogeological) methods are used to estimate models of spatially distributed subsurface properties. In hydrogeology, two paths to obtain hydrologically relevant subsurface models are by means of tomographic methods that are based solely on hydraulic signals such as hydraulic tomography (Butler et al., 1999; Yeh and Liu, 2000; Brauchler et al., 2011) or by approaches making use of the generally higher resolving power of geophysical methods (Hyndman et al., 1994; Linde et al., 2006; Hinnell et al., 2010; Lochbühler et al., 2013). Such methods are useful for detailed characterization on a local scale, but it is also important for the hydrogeologist to have an idea about the general hydrogeological setting and subsurface heterogeneity. Such knowledge can then be used to complement sparsely distributed measurements of hydraulic conductivity, or other properties of interest, by means of geostatistical simulation (Kitanidis, 1997), which in turn enables flow and transport predictions at larger scales. It is often more relevant to use geophysical measurements to compare conceptual models of subsurface heterogeneity than to improve the estimates of individual model cells (Linde, 2014).

Lochbühler et al. (2014) present a workflow to condition geostatistical simulations to tomograms and illustrate the method for two synthetic case studies. The resulting realizations honor the geophysical information present in a tomogram together with the spatial correlations between model cells described by a geostatistical model that is assumed known. This allows representing sub-resolution features that are included in the geostatistical model but not resolved by the geophysical inversion. Furthermore, multiple uncorrelated model realizations can be generated. The method avoids the creation of long Markov chain Monte Carlo (MCMC) walks as required in purely probabilistic inversion (e.g., Mosegaard and Tarantola, 1995). Nevertheless, note that the conditioning is made with respect to the geophysical tomograms, not to the actual underlying geophysical data, which would be the case for MCMC. Lochbühler et al. (2014) propose a post-processing step in which the geophysical forward response of the conditional realizations is evaluated and only those that fit the observed data are kept for further analysis.

A finding that was not reported by Lochbühler et al. (2014) is that inadequate geostatistical descriptions of heterogeneity produce conditional realizations with a corresponding geophysical forward response (i.e., simulated travel times) with a significantly higher data misfit than when the correct geostatistics are used. This suggests that the method could potentially be used to compare and rank different descriptions of subsurface heterogeneity (i.e., different geostatistical models) in terms of the geophysical data misfit of the conditioned realizations. An alternative approach was recently presented by Hermans et al. (2015) that investigate to what extent electrical resistivity data can be used to falsify or support alternative conceptual geological models of an alluvial aquifer in Belgium. Following the approach by Park et al. (2013), they do not seek local conditioning, but agreement in terms of global patterns and statistics.

The MAcro-Dispersion Experiment (MADE) site at the Columbus Air Force Base, Mississippi, USA, is a well-studied hydrogeological research site. It is of particular interest, as many different attempts have been made to describe the highly heterogeneous sediments forming the aquifer body and to understand anomalous flow- and transport observations (Zheng et al., 2011). Relevant studies concerning the MADE site include the site description by Boggs et al. (1992), results from an early tracer experiment revealing a non-Gaussian shape of the plume (Adams and Gelhar, 1992) and a geostatistical description of the hydraulic conductivity using a detrended exponential variogram (Rehfeldt et al., 1992). Numerous alternative approaches to describe the heterogeneity in hydraulic conductivity and to correctly reproduce the observed tracer behavior have been proposed over the last two decades (Eggleston and Rojstaczer, 1998; Harvey and Gorelick, 2000; Feehley et al., 2000; Benson et al., 2002; Barlebo et al., 2004; Salamon et al., 2007; Ronayne et al., 2010; Bohling et al., 2012; Meerschaert et al., 2013). A common finding in these works is that reproducing the plume's asymmetry requires an adequate description of the local-scale hydraulic conductivity (Eggleston and Rojstaczer, 1998; Harvey and Gorelick, 2000; Feehley et al., 2000; Dogan et al., 2014).

There is a broad consensus that stationary multivariate Gaussian (multi-Gaussian) descriptions of spatial heterogeneity (i.e., in form of a continuous variogram) are severely limited in producing geostatistical realizations with long range connectivity of high or low hydraulic conductivities (Silliman and Wright, 1988; Rubin and Journel, 1991; Gómez-Hernández and Wen, 1998). In such models, extreme values are spatially uncorrelated by construction. To achieve higher degrees of connectivity, truncated Gaussian and pluri-Gaussian models have been introduced, in which discrete facies are obtained by applying thresholds to simulated multi-Gaussian fields (Allard, 1994; Mariethoz et al., 2009). Zinn and Harvey (2003) propose a method to transform multi-Gaussian fields into fields with Gaussian statistics but where high- or low-conductivity values are well connected. No truncation is needed in their approach and the Gaussian univariate connectivity distribution is preserved. It has also been suggested that formulations based on fractal distributions provide better predictions than those based on classical Gaussian assumptions (Meerschaert et al., 2013).

As an alternative to variogram-based models, multiple-point statistics (MPS) describe the spatial dependencies between sets of points larger than 2. The correlation structure is not described in parametric form, but rather through a training image (TI) that contains the expected lithological units and their shapes and patterns (Strebelle, 2002; Hu and Chugunova, 2008; Mariethoz and Caers, 2008). MPS simulations are capable of generating discrete property fields with curvilinear, elongated and/or repeating structural elements. Note that MPS simulations can still generate multi-Gaussian fields, a characteristic we will use herein.

The potential of geophysical methods to obtain information about the spatial correlation of subsurface properties has been demonstrated, for instance, by Irving et al. (2010) who infer the aspect ratio of horizontal to vertical correlation length from radar and seismic reflection data. However, this is only feasible when carefully accounting for the influence of resolution limitations in the interpreted sections or tomograms (Day-Lewis et al., 2005; Moysey et al., 2005). In this work, we test if different descriptions of heterogeneity can be ranked by means of their tomographic response. This in turn would allow decreasing the space of possible conceptual models that are in agreement with field data as incompatible model types can be removed from further analysis. Here, we use crosshole ground-penetrating radar (GPR) first-arrival travel times to create a 2-D tomographic image that are used to condition geostatistical realizations of hydraulic conductivity. We first describe the geostatistical models of MADE that we considered in this study and how geostatistical realizations are generated. Next, the conditioning approach and the petrophysical links used to relate the hydraulic conductivity to GPR-relevant properties are introduced. We show results of the conditioning and how the conditional realizations for different geostatistical models behave in terms of their geophysical data predictions. Finally, the results are discussed against the background of previous studies at the MADE site and the main conclusions of the present work are drawn.

**2. A motivating example**

A synthetic motivating example is first presented before we consider the geostatistical models and the data specific to the MADE site. We leave the detailed description of the methodology to the following sections. Consider two reference models that represent the actual subsurface heterogeneity at two imaginary sites: one comprising channels (reference 1; Figure 1a) and the other comprising lenses (reference 2; Figure 1d). For each of these reference models we calculate a synthetic tomogram (not shown; see Lochbühler et al. (2014) for examples) based on noise-contaminated forward-simulated data and condition geostatistical simulations of hydraulic conductivity to these tomograms using the method of Lochbühler et al. (2014). The assumed geostatistical characteristics are in both cases represented by a training image featuring channels (i.e., in agreement with the first reference model). This leads to conditional realizations of hydraulic conductivity that include channel structures as enforced by the assumed geostatistical model and they generally reproduce the locations of high- and low porosity zones in the reference models (Figures 1b and e). They explain the data for reference 1 (Figure 1d), but not for reference 2 as the realizations fail to predict the measured data satisfactorily. Large geophysical data misfits of the conditional realizations can thus be interpreted as an indicator of an inadequate conceptual model as represented by the training image.

## 3. Geostatistical models

In this section we describe six different alternative conceptual models of the MADE site that are to be compared. These models only provide a small subset of all possible conceptual models and they are primarily introduced to demonstrate a methodology than to make definite statements about the MADE site.

*3.1 Multi-Gaussian Fields*

Our first conceptual model describes the heterogeneity in the natural logarithm of hydraulic conductivity, $Y=\ln(K)$, at the MADE site as a traditional multi-Gaussian field of exponential type defined by a mean log hydraulic conductivity $\mu_{\ln K} = -0.2627$ m/d, a ln$K$ variance of 6.6 and correlation lengths of 10 and 1 m in the $x$- and $z$-direction, respectively. These values are inspired by Bohling et al. (2012) who performed a geostatistical analysis of more than 30,000 direct-push injection logger (DPIL) hydraulic conductivity measurements. Multiple realizations with the prescribed properties were generated by multiplication of the characteristic covariance matrix **C** with random Gaussian fields (Alabert, 1987). The discretization for these and all the following geostatistical model realizations is 0.05 m. Several unconditional realizations of hydraulic conductivity are shown in Figure 2a.

*3.2 Disconnected and connected fields*

In multi-Gaussian fields, the conductivity values that are close to the mean have the highest connectivity. To investigate fields that have the same histogram and covariance as the multi-Gaussian fields (section 3.1) but different connectivity patterns, we follow the approach by Zinn and Harvey (2003). In this procedure, multi-Gaussian fields are transformed into fields where high- or low values are strongly connected, while the underlying spatial statistics (mean, variance, correlation lengths) are preserved. First, the absolute values of the original multi-Gaussian field are transformed into a zero-mean unit-variance field, in which extreme values become high values and values close to the mean become low values (Zinn and Harvey, 2003; Renard and Allard, 2013). Using a normal score transform, the resulting histogram is retransformed into a Gaussian distribution

$$Y' = \sqrt{2}\,\text{erf}^{-1}\left(2\text{erf}\left(\frac{Y}{\sqrt{2}}\right) - 1\right). \tag{1}$$

In the field *Y'*, low hydraulic conductivities are well connected. The so-derived conductivity values are then mirrored around their mean to obtain another field where high-conductivity values are well connected. Since the correlation lengths are reduced by the absolute transform, we define a rescaling factor by calculating the autocorrelation in the *x*- and *z*-direction for certain lags and compare these correlations to those of the original multi-Gaussian field. The discretization is then adjusted such that the final connected field has the same correlation structure (and mean and variance) as the original field. Following Zinn and Harvey (2003), we refer to fields where the high-conductivity values are well connected as 'connected', and fields where the low-conductivity values are well connected as 'disconnected'. Examples of disconnected and connected fields are depicted in Figures 2b and 2c, respectively. Visually, the realizations appear to show the opposite behavior; this is due to the fact that the correlation length in the *x*-direction (10 m) is larger than the borehole separation (4.25 m).

*3.3 Hybrid multi-Gaussian/multiple-point statistics fields*

Multiple-point statistics (MPS) describe the spatial dependencies of subsurface properties by statistics of orders higher than 2. MPS has been developed to overcome limitations of variogram-based models (Strebelle, 2002; Hu and Chugunova, 2008). The higher order dependencies are typically expressed in form of a training image, that is, an image that contains the dominant geological facies and the expected structural patterns. Training images can be based on outcrop data, on expertise concerning the expected stratigraphy and dominant geologic processes, or on logging or other auxiliary information.

Ronayne et al. (2010) proposed a description of subsurface property heterogeneity at the MADE site that combines continuous multi-Gaussian fields and MPS simulations. Based on extensive flowmeter data of hydraulic conductivity (Rehfeldt et al., 1992; Salamon et al., 2007), they modeled multi-Gaussian background fields, where the correlation structure is described by an anisotropic spherical variogram of the form (Salamon et al., 2007)

$$\gamma(\mathbf{h}) = \begin{cases} c_0 + c_1 \left[ 1.5 \left( \frac{|\mathbf{h}|}{a} \right)^3 \right], & \text{if } |\mathbf{h}| \leq a \\ c_0 + c_1, & \text{if } |\mathbf{h}| \geq a \end{cases} \quad (2)$$

where $c_0$ and $c_1$ depict the nugget and the sill, respectively, $a$ is the range and $\mathbf{h}$ the separation vector. These background fields are generated by multiplication of the characteristic covariance matrix with Gaussian random noise, similar to the multi-Gaussian fields (see above). It has been argued that the non-Fickian behavior of tracer transport at the MADE site may be related to high-conductivity channels formed by coarse open-framework gravel units observed in core samples (Ronayne et al., 2010; Bianchi et al., 2011a), but modeling efforts using the preferential flowpath approach have so far not been universally successful at reproducing observed plume behavior at the MADE site. The coarse open-framework gravel lithofacies has also been observed in a nearby outcrop and is part of the training images described below. The high-K channels are here simulated by multiple-point direct sampling of a binary training image featuring connected channels in a homogeneous matrix (Mariethoz et al., 2010b). The original training image by Strebelle (2002) shown in Figure 3a was adjusted such that the channel width, the channel fraction and the hydraulic conductivity (constant at 250 m/d) are in agreement with the channels modeled by Ronayne et al. (2010). The resulting hybrid multi-Gaussian/MPS realizations were then created by overlaying the binary MPS realizations on the multi-Gaussian fields (Figure 2d).

*3.4 Outcrop-based training image realizations*

The MADE site has been the object of extensive studies, including the description of hydrogeological facies and sedimentary units and their dominant patterns (Rehfeldt et al., 1992; Dogan et al., 2011). For this study, we created a training image based on hydrogeological facies mapping of a nearby outcrop (Rehfeldt et al., 1992). This outcrop is too small to form a training image. To create a training image of appropriate size, we proceeded by placing pieces from the outcrop at random into a large empty grid. Multiple-point direct sampling (Mariethoz et al., 2010) was then applied to fill the empty spaces between these pieces by simulation (see the resulting training image in Figure 3b) to obtain similar patterns as those found in the outcrop. To obtain the large training image, patterns from the outcrop are used to fill the undefined cells. Each facies was assigned the hydraulic conductivity value observed by Rehfeldt et al. (1992) (Table 1). Unconditioned realizations (Figure 2e) of hydraulic conductivity are then generated by direct sampling of the training image.

*3.5 Analog-based training image realizations*

Additional to the outcrop-based training image described in the previous section, we use a training image that is chosen purely based on the knowledge of the sedimentary environment at the MADE site, that is, the aquifer is formed by alluvial terrace deposits with different sand- and gravel units (Boggs et al., 1992). A training image featuring typical sedimentary structures of alluvial deposits is available from a detailed 3-D mapping study at the Herten site in SW Germany (Bayer et al., 2011; Comunian et al., 2011). Here we test to what extent geostatistical realizations based on the Herten model can be conditioned to the GPR tomogram from the MADE site. Following Lochbühler et al. (2014), the available 3-D TI was reduced by only considering a 2-D slice and the ten observed facies at the Herten site were reduced to the four facies observed in the MADE outcrop (Rehfeldt et al., 1992). The geological units in the TI are thus the same as for the outcrop-based TI. The dominant structural elements are gravel sheets, erosional surfaces and cross bedding. See Figure 3c for the TI and Figure 2f for individual unconditioned realizations of hydraulic conductivity.

## 4. Generation of conditional hydraulic conductivity models

*4.1 Data and original tomogram*

The crosshole GPR travel time data were acquired between two boreholes in the MultiLevel Sampler (MLS) cube with borehole separation of 4.25 m. We refer to Bianchi et al. (2011b), Dogan et al. (2011), and Bohling et al. (2012) for detailed descriptions of the site and the borehole locations. First-arrival travel times were picked manually and the picks were refined automatically using a statistically based information content picker (AIC picker, Leonard, 2000). Transmitter and receiver station spacing was 0.25 m. We only considered travel times with ray angles smaller than 50° inclination from the horizon, resulting in a data set of 974 first arrivals. The measurement errors are assumed to follow a zero-mean and uncorrelated Gaussian error model with a standard deviation of 1.4 ns due to picking and geometrical errors.

Inversion is performed by smoothness-constrained least-squares fitting of the travel time data (Lochbühler et al., 2014). To regularize the inverse problem, we applied an anisotropic first-order difference roughness operator with a horizontal-to-vertical anisotropy ratio of 10:1. This ratio corresponds to the anisotropy observed by Bohling et al. (2012) and is consistent with our geostatistical models. We consider the radar signal to propagate along curved ray paths between the transmitter and receiver positions, where the ray paths depend on the GPR velocity field. The inverse problem is thus non-linear and is solved iteratively by subsequently updating the model until the measured data are fitted according to the error model. The forward solver solves the Eikonal equation using the finite-difference algorithm by Podvin and Lecomte (1991) and ray-tracing is performed for each receiver location (Vidale, 1988).

The resulting tomogram features several high- and low-velocity regions that are well defined despite the generally smooth image (Figure 4). The inherent smoothness in the tomogram limits its usefulness for geological interpretation, which is a general limitation of smoothness-constrained deterministic inversion (e.g., Ellis and Oldenburg, 1994). The objective of our geostatistical conditioning approach that accounts for the resolution limitations of geophysical tomograms is to partly overcome this shortcoming.

*4.2. Conditioning procedure*

The approach presented by Lochbühler et al. (2014) allows us to condition geostatistical simulations to geophysical tomograms. We briefly describe the conditioning workflow here; for a detailed description we refer to Lochbühler et al. (2014). For each of the geostatistical scenarios described above, we generate 1000 unconditioned realizations of the ln$K$ field (i.e., similar to the realizations shown in Figure 2). All these realizations are then subject to synthetic geophysical forward and inverse modeling. We simulate crosshole GPR experiments, mimicking the true experiment conducted in the field and invert the so created data to obtain a GPR tomogram (i.e., a GPR velocity model) for each realization. Inversion is performed by smoothness-constrained least-squares fitting of the data, thereby mimicking the procedure applied to the field data (see previous section).

The 1000 realizations and the corresponding tomograms form a bivariate training image that contains the expected subsurface properties and their spatial distribution as one variable, and the corresponding geophysical tomogram as another variable. This training image is then sampled for patterns found in the tomogram (Figure 4) using the multiple-point direct sampling algorithm *DeeSse*, which is a commercial version of the original direct sampling algorithm by Mariethoz et al. (2010). The tomogram obtained by inversion of the real data from the MADE site is thus used as a conditioning image. The pattern sampling mechanism of the direct sampling algorithm for a bivariate TI is illustrated in Figure 5. A pattern is defined by the *n* pixels that are the closest to the pixel to be simulated that already have an assigned value in the simulation grid, for example, as exemplified by the three pixels (light green, green and light blue) in the 'Conditioned realization' in Figure 5. The pattern is built by the pixel values of the *n* cells and the lag vectors that denote their location relative to the pixel of interest. The bivariate TI is then scanned until a layer in the bivariate TI is found for which (a) the distance between the pattern projected on the original tomogram and the pattern found in the tomogram is below a threshold $t_{tomogram}$ and (b) the distance between the pattern in the simulation grid and the pattern found in the ln*K* realization is below a threshold $t_{\ln K}$. Once such a layer is found (as the bottom layer in Figure 5), the pixel value from the corresponding ln*K* realization is pasted into the simulation grid (the dark red pixel in Figure 5). For continuous variables, the pattern distance is calculated using the $l_1$ or $l_2$ norm distance. For categorical variables, we use the sum of non-matching cells (out of the *n* cells) as distance measure. To ensure a reasonable sampling efficiency, if a certain fraction *f* of layers in the bivariate TI is scanned and no acceptable match is found, the match with the smallest distance is used. The algorithmic parameters are depicted in Table 2. Mariethoz et al. (2010) and Meerschman et al. (2013) provide further details on the direct sampling algorithm and provide guidance concerning the choice of algorithmic variables. Note that the TI is scanned 'vertically', meaning that the pattern is not moved within the plane of the realizations but the pattern is projected onto individual layers until a match is found. This is done to account for resolution variations within the tomogram (see Lochbühler et al. (2014) for details). The resulting geostatistical realizations honor the geophysical tomogram (i.e., a slowness field in this study), but there is no guarantee that they also honor the geophysical data (first arrival travel times in this study). Lochbühler et al. (2014) proposed a post-processing step in which the geophysical forward response of each realization was evaluated against the observed geophysical data.

*4.3. Petrophysics*

The conditioning method involves calculating synthetic geophysical responses for the set of unconditional realizations. A petrophysical relationship is thus needed to translate the hydraulic conductivity fields into fields of the geophysical target property, which for the crosshole GPR travel times used herein is the radar slowness *u* (i.e., the reciprocal of the velocity).

The relation between radar slowness and porosity can be described by (Pride, 1994; Davis and Annan, 1989)

$$u = \frac{1}{c}\sqrt{\varphi^m \left(\varepsilon_w - \varepsilon_s\right) + \varepsilon_s}, \tag{3}$$

where $\varphi$ is the porosity, *c* is the speed of light in a vacuum, $m[-]$ denotes the cementation factor, and $\varepsilon_w$ [ ] and $\varepsilon_s$ [ ] are the relative electrical permittivity of water and solid grains, respectively. Here, we use $c = 3 \times 10^8$ m/s, $\varepsilon_w = 81$, $\varepsilon_s = 3$ and $m = 1.65$. The cementation factor $m$ was tuned using the tomogram from MADE (Figure 4) to obtain the mean porosity of 0.3 that was observed by Boggs et al. (1992). Note that Adams and Gelhar (1992) suggested a value of 0.35, which highlights a certain amount of uncertainty on the proper value to choose. Relating porosity to hydraulic conductivity values is challenging due to strong site and lithology dependencies. Based on equation (3), we plotted the tomogram-derived porosity at the transmitter borehole (from the original tomogram) against a cm-scale direct-push log of hydraulic conductivity by Bohling et al. (2012) (Figure 6a). Regression analysis of the two properties yields a linear relation of the form

$$\varphi = 0.0055 \ln(K) [\text{m}/\text{d}] \quad (R = 0.58) \tag{4}$$

To match the different resolutions in the geophysically inferred porosity and the direct push data, the log of hydraulic conductivity was smoothed prior to regression with a Gaussian window of a length of 1 m (see scatter plot in Figure 6b). Finding a valid petrophysical link is a crucial step in the workflow and the present work is somewhat limited by the lack of detailed site-specific information on porosity or permittivity. However, we argue that the petrophysical relationship used herein is justified for comparison purposes. The same error is made for all geostatistical scenarios, so it is likely that a scenario that does better than the others is more appropriate.

## 5. Comparison of alternative geostatistical models

For all of the considered geostatistical models (section 3), examples of tomogram-conditioned realizations are shown in Figure 7 (1000 conditional realizations were created for each type of geostatistical model). The realizations reproduce the large-scale property distribution observed in the original tomogram, but each geostatistical model produces realizations with different types of small-scale heterogeneities that are in agreement with the respective training images (Figure 2). The low velocity zone in the center of the model domain detected by the GPR measurements is in most cases translated into a distinct high hydraulic conductivity zone. The outcrop-based MPS realizations (Figure 7e) do not feature this extended zone, likely due to an underrepresentation of high *K* values in the corresponding TI and the smaller range of *K* values reported by Rehfeldt et al. (1992) compared with Bohling et al. (2012). In the hybrid realizations (Figure 7d), the large high-K zone in the center is based on the multi-Gaussian background field, whereas the small-scale fluctuations of high conductivity are represented by the discrete high-K channels from the MPS description.

Next, the hydraulic conductivity predictions are compared to the available high-resolution *K*-log at the transmitter borehole (Figure 8). All geostatistical model realizations underestimate the hydraulic conductivity in the upper part of the aquifer. This happens as the petrophysical relation (eq. 4) performs rather badly in this depth interval (see Figure 6a). The multi-Gaussian (Figure 8a) and disconnected (Figure 8b) realizations overestimate the hydraulic conductivity in the central depth section. The outcrop-based MPS simulations underestimate the hydraulic conductivity along the entire depth profile, indicating that the high *K* facies are underrepresented in the TI (Figure 8e). This implies that the geostatistical characteristics of the local outcrop offer a poor description of the local heterogeneity around the MLS cube. This result is perhaps not surprising given the extreme heterogeneity of the MADE site (Bohling et al., 2012) and it calls for constructing larger-scale training images that represent all types of heterogeneity that can be encountered at the MADE site. It happens that the analog-based MPS realizations (Figure 8f) capture the measured *K* values quite well. Overall, abrupt changes in the *K*-log are accompanied by high gradients in the *K* predictions, which is a result of the strong dependency on the applied petrophysical relation.

Another way to assess the different geostatistical models is by comparing the data predictions of the model realizations (Figure 9). The data fit is expressed as a weighted root mean square error

$$WRMSE = \sqrt{\frac{1}{N}\sum_{j=1}^{N}\frac{\left(d_j - F_j(m)\right)^2}{\sigma_j^2}} \qquad (5)$$

where *N* is the number of GPR travel times, $d_j$ and $F_j(\mathbf{m})$ are the measured and predicted travel times of observation $j$, **m** is the tested model realization and $\sigma_j$ (1.4 ns) describes the corresponding measurement and modeling error. A WRMSE of 1 thus means that the travel time data are on average fitted to the measurement errors.

The density distributions of the data predictions of the 1000 unconditioned realizations of each geostatistical model type are depicted in Figure 9a. None of these realizations fit the data and the spread is wide. After conditioning (Figure 9b), the distributions are centered at lower values and the spread has decreased, especially for the continuous and hybrid fields. The multi-Gaussian, disconnected and hybrid realizations predict the measured data reasonably well (WRMSE<1.2). The data fits of connected and MPS (analog-based) realizations are slightly worse, and the spread of the data predictions is larger. For the outcrop-based MPS realizations, the width and the mean of the data fit distribution remains almost unchanged compared to the unconditional realizations, strongly indicating that the corresponding training image does not represent the subsurface adequately at this specific location of the MADE site.

## 6. Comparison of alternative petrophysical models

Using the analog-based MPS model as an example, we demonstrate that alternative petrophysical models can be readily tested in the presented framework. The tomogram-conditioned realizations (Figure 10a) based on equation 4 (hereafter referred to as petrophysical model 1) are in agreement with the observed data (Figure 11) and is dominated by high *K* lithofacies.

Negative correlations between hydraulic conductivity and porosity have been reported for similar geological settings (Morin, 2006). They are expected when fine sediments (high porosity and low hydraulic conductivity) constitute a significant portion of the aquifer material. We re-ran the conditioning scheme using an opposite sign on the slope in equation 4 (petrophysical model 2). By doing so, we preserve the porosity range and the mean porosity observed by Boggs et al. (1992), as well as the velocity range in the original tomogram. The resulting conditional realizations are shown in Figure 10b. The best conditioning is achieved by describing the conductivity fields as predominantly homogeneous, but the data misfits are high (Figure 11).

We also considered a case in which the TI has been built in terms of hydrofacies without direct porosity information. For this scenario, we consider the extensive review by Heinz et al. (2003) that provides porosity and hydraulic conductivity values of different facies found in glaciofluvial gravel bodies. From there, we extract discrete porosity values for the conductivities of the individual hydrofacies (see Table 1). To produce fields where the GPR velocity values are in the range observed in the original tomogram, the cementation factor was set to $m$ = 1.35. This implies that the cementation factor was used as a tuning parameter to relate the porosities presented by Heinz et al. (2003) to the observed GPR velocities. This scenario is referred to as petrophysical model 3. The conditional realizations display highly variable hydraulic conductivity fields that mainly feature facies of moderate and high hydraulic conductivities (Figure 10c). The corresponding data misfits are very high (Figure 11).

Based on the simulated data predictions, both petrophysical models 2 and 3, which are not based on local information, are found to be unsuitable to describe the petrophysics for this location of the MADE site. This conclusion is expected as any significant relationship between hydraulic conductivity and porosity (and hence GPR velocity) is expected to be site-specific (e.g., Morin, 2006; Purvance and Andricevic, 2000). However, this simple comparison suggest not only that reliable site-specific relationships are needed, but also that it is possible to identify poor petrophysical relationships by comparing the forward response of the conditional realizations with the observed data (Figure 11). These findings are also representative for the other geostatistical scenarios (not shown) tested within the scope of this study.

## 7. Model selection

Quantitative classification of competing conceptual model formulations, or model types, is the premise of model selection. Model types can comprise different choices of the model parameterization, the underlying physical relations or error models. Any choice of how the physical system is represented in the inverse problem corresponds to a specific model type (Dettmer et al., 2009). Here, the considered model types are defined by the different geostatistical (section 3) and petrophysical (section 6) models.

Model selection is often embedded in a Bayesian framework, such that model types can be compared by their probabilistic relevance, or the selection process between two competing model types is formulated as a hypothesis test. For the latter, a characteristic and easily quantifiable hypothesis is tested against its counterhypothesis by counting the number of occurrences of the hypothesis being true and false in a set of model realizations (Khan and Mosegaard, 2002). If the model types to be compared cannot be characterized by a single criterion to be fulfilled or not, their relevance can be expressed by their evidence. In a formal Bayesian sense, the evidence $E_i$ of a model type $M_i$ is the probability that the measured data **d** results from the model type $M_i$ and it is given by

$$E_i = p(\mathbf{d}|M_i) = \int p(\mathbf{d}|\mathbf{m}_i, M_i) p(\mathbf{m}_i|M_i) d\mathbf{m}_i, \tag{6}$$

where $p(\mathbf{d}|\mathbf{m}_i, M_i)$ is the probability that the measured data is predicted by the individual realization $\mathbf{m}_i$ from the model type $M_i$, that is, the likelihood of the realization $\mathbf{m}_i$. The prior probability $p(\mathbf{m}_i|M_i)$ describes the probability that $\mathbf{m}_i$ is a realization of the model type $M_i$. As the integral in Equation 6 spans over the entire parameter space, $E_i$ is the marginal probability of the data (Kass and Raftery, 1995).

Forming the ratio between the evidences of two competing model types yields the Bayes factor

$$B_{ij} = \frac{p(\mathbf{d}|M_i)}{p(\mathbf{d}|M_j)}, \tag{7}$$

which provides a quantitative measure for the favor of one model type $M_i$ over model type $M_j$ (Jeffreys, 1961).

In practice it is challenging to obtain a robust estimate of the evidence, as the integrand in Equation 6 is potentially multi-modal and highly peaked. For cases where sampling is prohibitive, asymptotic approximations to estimate the evidence can be used. Assuming that the information content in the prior probability is marginal and the evidence is well characterized by the maximum likelihood estimate $\hat{L}$, the model type relevance can be expressed by the Bayesian Information Criterion (BIC, Schwarz, 1978; Kass and Raftery, 1995):

$$BIC_i = -2\log(\hat{L}) + k\log(N), \qquad (8)$$

where $\hat{L}$ is the maximum value of the likelihood function, $k$ is the number of model parameters and $N$ is the number of data. Note that lower BIC values denote higher model relevance and model parsimony is enforced as simple models (low $k$) are given higher relevance.

As the inverse problem was not solved in a Bayesian sense, no formal probability density functions are available. However, the unconditional realizations provide a set of prior realizations as they are generated solely based on the underlying geostatistics without considering any data. Similarly, the conditioned realizations can be interpreted as a sort of posterior realizations, although it must be stressed that unlike formal posterior realizations, these are not conditioned to measured data through a likelihood function, but only to the original tomogram. This implies that the forward response of the resulting conditional realizations do not necessarily fit the observed travel times. As each realization is conditioned to the tomogram individually, the obtained realizations are independent.

For each of the considered model types, we calculate the maximum a posteriori likelihood estimate $\hat{L}$ using

$$\hat{L} = \prod_{k=1}^{N} \frac{1}{\sqrt{2\pi\sigma_k^2}} \exp\left[-\frac{1}{2}\frac{\left(d_k^{\text{pred}}(\mathbf{m}) - d_k\right)^2}{\sigma_k^2}\right], \qquad (9)$$

from all the conditional realizations **m**. As the algorithmic parameters in the conditioning procedure remain unchanged, the direct sampling mechanism treats all models as of identical complexity. Additionally, the number of data does not change between the models, comparing the maximum likelihood estimates therefore amounts to comparing BIC values. The resulting estimates are shown in Table 3. Similar to the qualitative interpretation of the distribution of data predictions (Figures 9 and 11), we find that the analog-based MPS, Multi-Gaussian and the disconnected fields are the most likely, but we refrain (given the strong assumptions made) to differentiate between these three model types. Based on the significantly lower $\hat{L}$ values, we also suggest that the outcrop-based MPS model is inappropriate and that petrophysical models II and III can be discarded from further consideration.

## 8. Discussion

The geostatistical conditioning procedure used herein allows generating realizations for various geostatistical and petrophysical models, all conditioned to a single tomogram obtained by inversion of geophysical or hydrogeological data. The appropriateness of alternative model types can be addressed by analyzing different aspects of the conditional realizations.

The fit between the observed geophysical data and the data predictions of the individual model realizations is an obvious indicator to assess whether a model is potentially suitable to represent the true subsurface or not. As Madigan and Raftery (1994) put it, a model should not be considered if it predicts the data far less well than the model with the best predictions. Based on this decision rule, the outcrop-based MPS model and the alternative petrophysical models are clearly to be dismissed (Figures 9b and 11). Apart from comparing the model output to the measured geophysical data (Figure 9b) one can also compare the ln$K$ predictions of the conditional realizations to measured values, given that such information is available. We find that the multi-Gaussian, the disconnected and the analog-based MPS models show the best agreement with the ln$K$ data (Figure 8).

The maximum likelihood estimate (i.e., the data fit of the best-fit realization) proved a useful measure of model appropriateness in this study (Table 3). No distinct 'best' model type emerged from the study, but for two model types (multi-Gaussian and disconnected) the maximum likelihood estimate is significantly lower than for the others. An interesting finding of this study is that the outcrop-based MPS realizations using the petrophysical link derived from available log data predicts neither the GPR data nor the conductivity logging data satisfactorily. The high-hydraulic-conductivity zones (that are translated into low-velocity zones by the used petrophysical link) are underrepresented in the corresponding TI. This shows that local outcrops are not necessarily valid analogs for the subsurface and that creating a representative TI based on outcrop photographs or maps is not straightforward. The aquifer analog from Germany produced a more adequate training image (Figure 3c) as it included a sufficient proportion of high-$K$ lithofacies.

There is evidently a strong dependency of the conditioning outcome on the petrophysical link to translate the hydraulic properties (here, hydraulic conductivities) into the property sensed by geophysical measurements (here, GPR velocities). Finding an adequate petrophysical model is one of the major challenges for all approaches where hydraulic properties are to be inferred by geophysical experiments. As shown here, testing different petrophysical models is straightforward in the presented workflow and allows identifying inappropriate combinations of geostatistical models and petrophysical relationships. A possible way to circumvent the dependency on a potentially insufficient petrophysical relationship is to rely on hydraulic rather than geophysical tomography where the measured data are directly sensitive to hydraulic conductivity.

Implications on the subsurface heterogeneity at the MADE site are that (a) if the porosity-conductivity trend inferred from the available $K$-log can be relied on, high conductivity zones are required to explain the observed GPR data; (b) multi-Gaussian realizations and derivatives thereof (disconnected fields based on the methodology by Zinn and Harvey (2003)) can be successfully conditioned to the field-based tomogram, and to obtain realizations that explain the geophysical data without the need for discrete facies boundaries; (c) an outcrop from Germany provided a better training image than the one based on a local outcrop from Rehfeldt et al. (1992). These findings about the MADE site are only based on a subset of alternative conceptual and petrophysical models with conditioning to a resolution-limited tomogram. Extending this work to include higher-resolution surface GPR, tracer test data, and other conceptual models of the $K$ field, is left for the future. We also postulate that it should be possible to build more appropriate and larger-scale training images of the MADE site than the one based on a very small and local outcrop. So far, we have only assimilated a small fraction of the thirty years of hydrogeological research that has been performed at this site (e.g., *Zheng et al.*, 2011) and there is ample room for improvements. It is likely that such an analysis would lead to more definite conclusions.

## 9. Conclusions

Direct sampling offers a powerful approach to condition multiple realizations of various underlying geostatistical models (represented in the form of a training image) to resolution-limited geophysical or hydrogeological tomograms. When the geostatistical model (i.e., training image) is poorly chosen, we demonstrate that the corresponding conditional realizations do not properly explain the geophysical data (crosshole GPR travel times in our study) that was used to construct the tomogram. This feature was used to demonstrate how to falsify and dismiss alternative conceptual models at the MADE site without assuming that one of the other considered conceptual models is "correct". We find that realizations based on a local outcrop are clearly inappropriate for representing the subsurface conditions at the considered location of the MADE site (the so-called MLS cube). We also found that petrophysical relationships that are not based on local information are unable to produce acceptable conditional subsurface realizations. Additional geophysical or hydrogeological models and the inclusion of additional sub-classes of different conceptual models are needed to better differentiate between the performance of multi-Gaussian fields (and its related derivatives), hybrid models and aquifer analogs.

Acknowledgements

This research was funded by the Swiss National Science Foundation (SNF) and is a contribution to the ENSEMBLE project (grant no. CRSI22 132249). Field work at the MADE site was supported by grants from the U.S. National Science Foundation (NSF, EAR-0738938 and EAR-0738955); any opinions, findings, and conclusions or recommendations expressed are those of the authors and do not necessarily reflect the views of the NSF. We thank James Butler and his colleagues from the Kansas Geological Survey for providing the logging data. We also thank Frederick Day-Lewis and two anonymous reviewers for their very constructive and insightful comments. This work is dedicated to the memory of PhD student Tobias Lochbühler who tragically lost his life in a mountaineering accident on 2014 July 19. Tobias was a truly wonderful person and a most gifted researcher.

**References**

Adams, E. E., Gelhar, L. W., 1992. Field study of dispersion in a heterogeneous aquifer: 2. Spatial moments analysis. Water Resources Research 28(12), 3293–3307.

Alabert, F., 1987. The practice of fast conditional simulations through the LU decomposition of the covariance matrix. Mathematical Geology 19(5), 369–386.

Allard, D., 1994. Simulating a geological lithofacies with respect to connectivity information using the truncated Gaussian model. In Geostatistical Simulations: Proceedings of the Geostatistical Simulation Workshop, pages 197–211.

Barlebo, H. C., Hill, M. C., Rosbjerg, D., 2004. Investigating the Macrodispersion Experiment (MADE) site in Columbus, Mississippi, using a three-dimensional inverse flow and transport model. Water Resources Research 40(4), W04211.

Bayer, P., Huggenberger, P., Renard, P., Comunian, A., 2011. Three-dimensional high resolution fluvio-glacial aquifer analog: Part 1: Field study. Journal of Hydrology 405, 1–9.

Benson, D. A., Schumer, R., Meerschaert, M. M., Wheatcraft, S. W., 2002. Fractional dispersion, Lévy motion, and the MADE tracer tests. In Dispersion in Heterogeneous Geological Formations, pages 211–240. Springer.

Bianchi, M., Zheng, C., Wilson, C., Tick, G. R., Liu, G., and Gorelick, S. M., 2011a. Spatial connectivity in a highly heterogeneous aquifer: From cores to preferential flow paths. Water Resources Research 47(5), W05524.

Bianchi, M., Zheng, C., Tick, G. R. and Gorelick, S. M., 2011b. Investigation of small-scale preferential flow with a forced-gradient tracer test. Groundwater 49, 503–514.

Boggs, J. M., Young, S. C., Beard, L. M., Gelhar, L. W., Rehfeldt, K. R., Adams, E. E., 1992. Field study of dispersion in a heterogeneous aquifer: 1. Overview and site description. Water Resources Research 28(12), 3281–3291.

Bohling, G. C., Liu, G., Knobbe, S. J., Reboulet, E. C., Hyndman, D. W., Dietrich, P., Butler, J. J., 2012. Geostatistical analysis of centimeter-scale hydraulic conductivity variations at the MADE site. Water Resources Research 48(2), W02525.

Brauchler, R., Hu, R., Dietrich, P., Sauter, M., 2011. A field assessment of high-resolution aquifer characterization based on hydraulic travel time and hydraulic attenuation tomography. Water Resources Research 47(3), W03503.

Butler, J., McElwee, C., Bohling, G., 1999. Pumping tests in networks of multilevel sampling wells: Motivation and methodology. Water Resources Research 35(11), 3553–3560.

Comunian, A., Renard, P., Straubhaar, J., Bayer, P., 2011. Three-dimensional high resolution fluvio-glacial aquifer analog: Part 2: Geostatistical modeling. Journal of Hydrology 405, 10–23.

Davis, J., Annan, A., 1989. Ground-penetrating radar for high-resolution mapping of soil and rock stratigraphy. Geophysical Prospecting 37(5), 531–551.

Day-Lewis, F. D., Singha, K., Binley, A. M., 2005, Applying petrophysical models to radar traveltime and electrical resistivity tomograms: Resolution-dependent limitations. Journal of Geophysical Research 110, B08206.

Dettmer, J., Dosso, S. E., Holland, C. W., 2009. Model selection and Bayesian inference for high-resolution seabed reflection inversion. The Journal of the Acoustical Society of America 125, 706–716.

Dogan, M., Van Dam, R. L., Bohling, G. C., Butler, J. J., Hyndman, D. W., 2011. Hydrostratigraphic analysis of the MADE site with full-resolution GPR and direct-push hydraulic profiling. Geophysical Research Letters 38(6), L06405.

Dogan, M., Van Dam, R. L., Liu, G., Meerschaert, M. M., Butler, J. J., Bohling, G. C., Benson, D., A., Hyndman, D. W., 2014. Predicting flow and transport in highly heterogeneous alluvial aquifers. Geophysical Research Letters 41, 7560-7565, doi:10.1002/2014GL061800.

Eggleston, J., Rojstaczer, S., 1998. Identification of large-scale hydraulic conductivity trends and the influence of trends on contaminant transport.

Water Resources Research 34(9), 2155–2168.

Ellis, R., Oldenburg, D., 1994. Applied geophysical inversion. Geophysical Journal International 116(1), 5–11.

Feehley, C. E., Zheng, C., Molz, F. J., 2000. A dual-domain mass transfer approach for modeling solute transport in heterogeneous aquifers: Application to the Macrodispersion Experiment (MADE) site. Water Resources Research 36(9), 2501–2515.

Gómez-Hernández, J., Wen, X., 1998. To be or not to be multi-Gaussian? A reflection on stochastic hydrogeology. Advances in Water Resources 21(1), 47–61.

Harvey, C., Gorelick, S. M., 2000. Rate-limited mass transfer or macrodispersion: Which dominates plume evolution at the Macrodispersion Experiment (MADE) site? Water Resources Research 36(3), 637–650.

Heinz, J., Kleineidam, S., Teutsch, G., Aigner, T., 2003. Heterogeneity patterns of quaternary glaciofluvial gravel bodies (SW-Germany): Application to hydrogeology. Sedimentary Geology 158(1-2), 1–23.

Hermans, T., Nguyen, F., Caers, J., 2015, Uncertainty in training image-based inversion of hydraulic head data constrained to ERT data: Workflow and case study. Water Resources Research, 51, 5332-5352.

Hinnell, A., Ferré, T., Vrugt, J., Huisman, J., Moysey, S., Rings, J., Kowalsky, M., 2010. Improved extraction of hydrologic information from geophysical data through coupled hydrogeophysical inversion. Water Resources Research 46(4), W00D40.

Hu, L., Chugunova, T., 2008. Multiple-point geostatistics for modeling subsurface heterogeneity: A comprehensive review. Water Resources Research 44(11), W11413.

Hyndman, D. W., Harris, J., Gorelick, S., 1994. Coupled seismic and tracer test inversion for aquifer property characterization. Water Resources Research 30(7), 1965–1978.

Irving, J., Scholer, M., Holliger, K., 2010. Inversion for the stochastic structure of subsurface velocity heterogeneity from surface-based geophysical reflection images. In Advances in Near-Surface Seismology and Ground-Penetrating

Radar, edited by R. Miller, J. Bradford, and K. Holliger, Society of Exploration Geophysicists, Tulsa, Oklahoma, 2010.

Jeffreys, H., 1961. Theory of Probability. Oxford University Press, Oxford UK.

Kass, R. E., Raftery, A. E., 1995. Bayes factors. Journal of the American Statistical Association 90(430), 773–795.

Khan, A., Mosegaard, K., 2002. An inquiry into the lunar interior: A nonlinear inversion of the Apollo lunar seismic data. Journal of Geophysical Research 107(E6), 5036.

Kitanidis, P., 1997. Introduction to geostatistics: Applications in hydrogeology. Cambridge University Press.

Leonard, M., 2000. Comparison of manual and automatic onset time picking. Bulletin of the Seismological Society of America 90(6), 1384–1390.

Linde, N. (2014), Falsification and corroboration of conceptual hydrological models using geophysical data. WIREs Water 1.

Linde, N., Finsterle, S., Hubbard, S., 2006. Inversion of tracer test data using tomographic constraints. Water Resources Research 42(4), W04410.

Lochbühler, T., Doetsch, J., Brauchler, R., Linde, N., 2013. Structure-coupled joint inversion of geophysical and hydrological data. Geophysics 73(3), ID1–ID14.

Lochbühler, T., Pirot, G., Straubhaar, J., Linde, N., 2014. Conditioning of multiple-point statistics facies simulations to tomographic images. Mathematical Geosciences 46(5), 625–645.

Madigan, D., Raftery, A. E., 1994. Model selection and accounting for model uncertainty in graphical models using Occam's window. Journal of the American Statistical Association 89(428), 1535–1546.

Mariethoz, G., Caers, J., 2015. Multiple-Point Geostatistics: Stochastic Modeling with Training Images. Wiley Blackwell, 364 pp.

Mariethoz, G., Renard, P., Cornaton, F., Jaquet, O., 2009. Truncated plurigaussian simulations to characterize aquifer heterogeneity. Ground Water 47(1), 13–24.

Mariethoz, G., Renard, P., Straubhaar, J., 2010. The direct sampling method to perform multiple-point geostatistical simulations. Water Resources Research

46(11), W11536.

Meerschaert, M. M., Dogan, M., Van Dam, R. L., Hyndman, D. W., Benson, D. A., 2013. Hydraulic conductivity fields: Gaussian or not? Water Resources Research 49(8), 4730–4737.

Meerschman, E., Pirot, G., Mariethoz, G., Straubhaar, J., Van Meirvenne, M., Renard, P., 2013. A practical guide to performing multiple-point statistical simulations with the direct sampling algorithm. Computers & Geosciences 52, 307–324.

Morin, R. H., 2006. Negative correlation between porosity and hydraulic conductivity in sand-and-gravel aquifers at Cape Cod, Massachusetts, USA. Journal of Hydrology 316(1), 43–52.

Mosegaard, K., Tarantola, A., 1995. Monte Carlo sampling of solutions to inverse problems. Journal of Geophysical Research 100(B7), 12431–12.

Moysey, S., Singha, K., Knight, R., 2005, A framework for inferring field-scale rock physics relationships through numerical simulation. Geophysical Research Letters 32, L08304.

Park, H., Scheidt, C., Fenwick, D., Boucher, A., Caers, J., 2013, History matching and uncertainty quantification of facies models with multiple geological interpretations. Computer and Geosciences 17, 609-621.

Podvin, P., Lecomte, I., 1991. Finite difference computation of traveltimes in very contrasted velocity models: A massively parallel approach and its associated tools. Geophysical Journal International 105(1), 271–284.

Purvance, D. T., Andricevic, R., 2000. On the electrical-hydraulic conductivity correlation in aquifers. Water Resources Research 36, 2905-2913.

Pride, S., 1994. Governing equations for the coupled electromagnetics and acoustics of porous media. Physical Review B 50(21), 15678.

Rehfeldt, K. R., Boggs, J. M., Gelhar, L. W., 1992. Field study of dispersion in a heterogeneous aquifer: 3. Geostatistical analysis of hydraulic conductivity. Water Resources Research 28(12), 3309–3324.

Renard, P., Allard, D., 2013. Connectivity metrics for subsurface flow and transport. Advances in Water Resources 51, 168–196.

Ronayne, M. J., Gorelick, S. M., Zheng, C., 2010. Geological modeling of submeter scale heterogeneity and its influence on tracer transport in a fluvial aquifer. Water Resources Research 46(10), W10519.

Rubin, Y., Journel, A. G., 1991. Simulation of non-gaussian space random functions for modeling transport in groundwater. Water Resources Research 27(7), 1711–1721.

Salamon, P., Fernàndez-Garcia, D., Gómez-Hernández, J., 2007. Modeling tracer transport at the MADE site: The importance of heterogeneity. Water Resources Research 43(8):W08404.

Schwarz, G., 1978. Estimating the dimension of a model. The annals of statistics 6(2), 461–464.

Silliman, S., Wright, A., 1988. Stochastic analysis of paths of high hydraulic conductivity in porous media. Water Resources Research 24(11), 1901–1910.

Strebelle, S., 2002. Conditional simulation of complex geological structures using multiple-point statistics. Mathematical Geology 34(1), 1–21.

Vidale, J., 1988. Finite-difference calculation of travel times. Bulletin of the Seismological Society of America 78(6), 2062–2076.

Yeh, T.-C. J. and Liu, S., 2000. Hydraulic tomography: Development of a new aquifer test method. Water Resources Research 36(8), 2095–2105.

Zheng, C., Bianchi, M., Gorelick, S. M., 2011. Lessons learned from 25 years of research at the MADE site. Groundwater 49(5), 649–662.

Zinn, B., Harvey, C., 2003. When good statistical models of aquifer heterogeneity go bad: A comparison of flow, dispersion, and mass transfer in connected and multivariate Gaussian hydraulic conductivity fields. Water Resources Research 39(3), 1051.

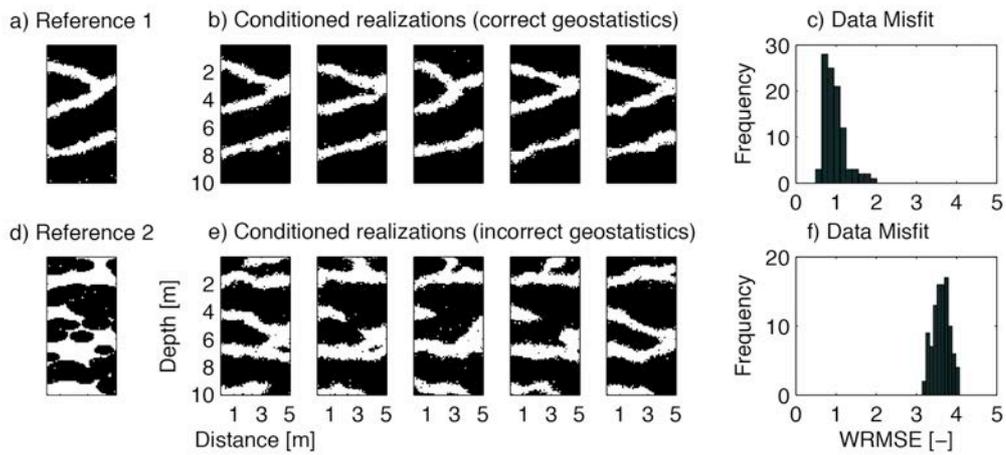

**Figure 1:** Illustration of how the MPS conditioning procedure can be used to compare alternative conceptual models. (a, d) Reference 'truths'; (b, e) geostatistical realizations conditioned to tomograms derived from the two references using a channel-based training image; (c, f) data misfit of 100 realizations, where WRMSE denotes the weighted root mean square error between measured and predicted GPR travel times (WRMSE = 1 indicates that the data are fitted to the expected error level). An inappropriate training image clearly leads to realizations that are not in agreement with the geophysical data.

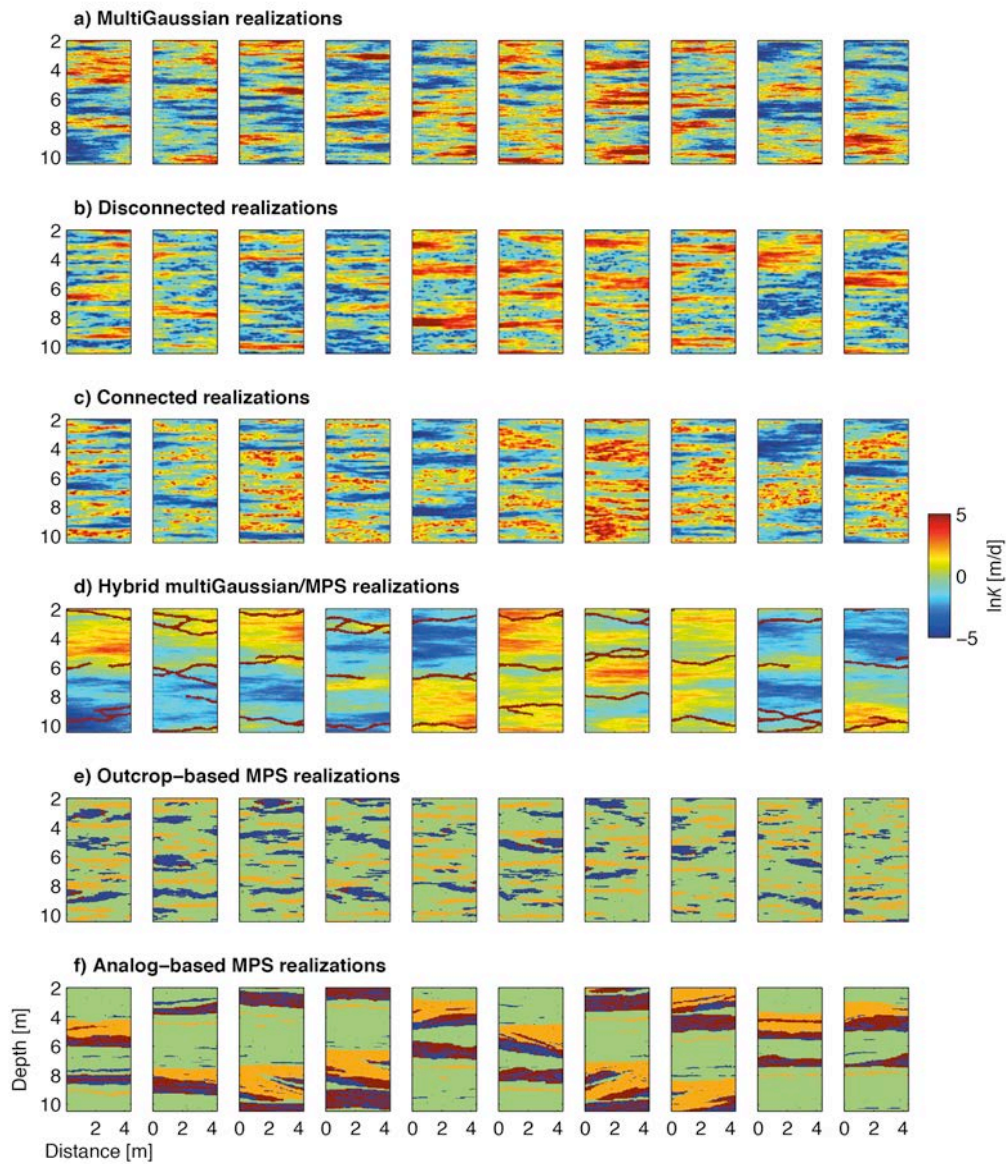

**Figure 2:** Unconditional realizations of the conceptual models considered at the MADE site. MPS is the abbreviation for multiple-point statistics.

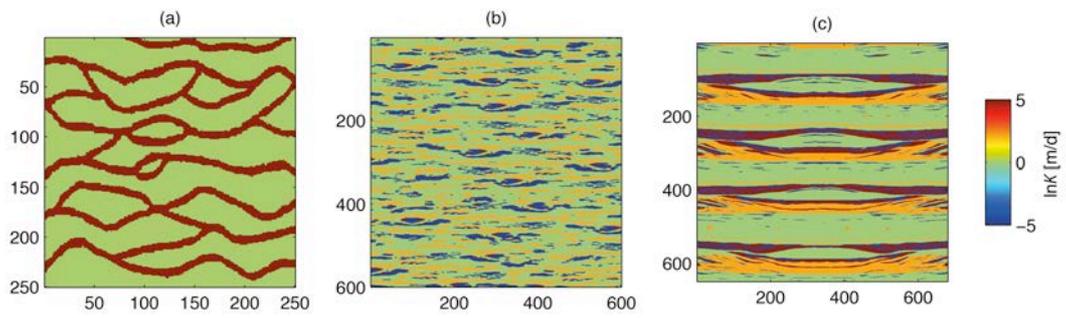

**Figure 3:** Training Images: (a) binary training image (Strebelle, 2002) used as basis for the simulation of continuous channels in the hybrid multi-Gaussian/MPS fields; (b) training image based on the outcrops of Rehfeldt et al. (1992); (c) training image based on the hydrogeological analog from the Herten site (Bayer et al., 2011; Comunian et al., 2011).

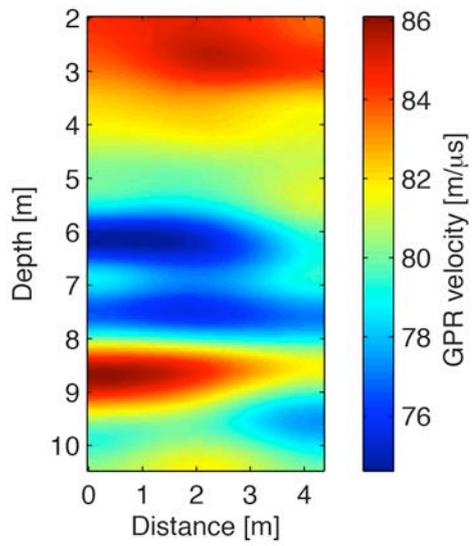

**Figure 4:** Original tomogram obtained from smoothness-constrained least-squares inversion of actual first-arrival GPR travel times at the MADE site.

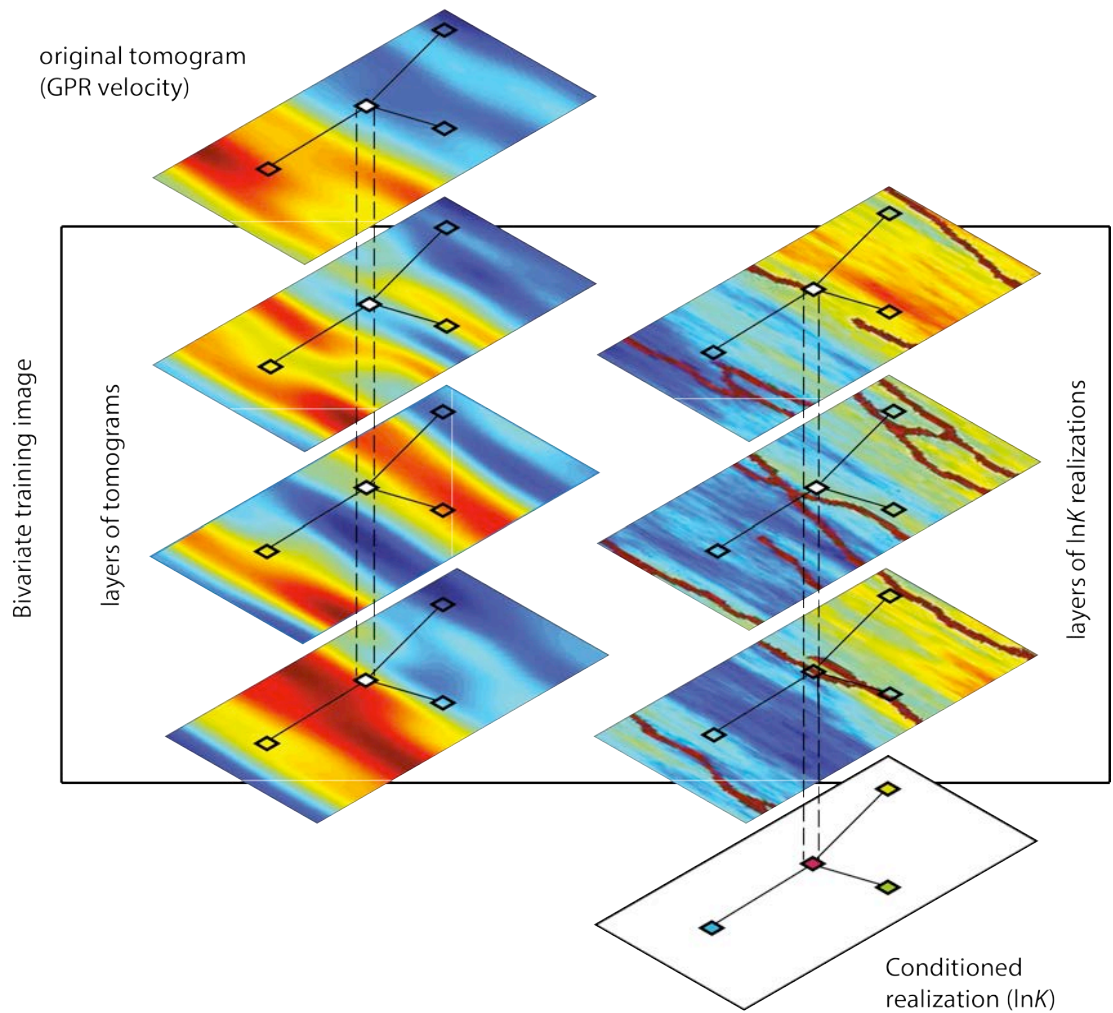

**Figure 5:** Conditioning procedure and sampling mechanism (modified after Lochbühler et al., 2014). A pattern is defined by the already defined cells in the conditional realization and projected on the original tomogram. The tomograms and hydraulic conductivity realizations are then scanned until the pattern matches the GPR velocities in the tomogram and the conductivity values in the hydraulic conductivity realization. Once a match is found, the cell value of interest found in the hydraulic conductivity realization is pasted into the conditional realization.

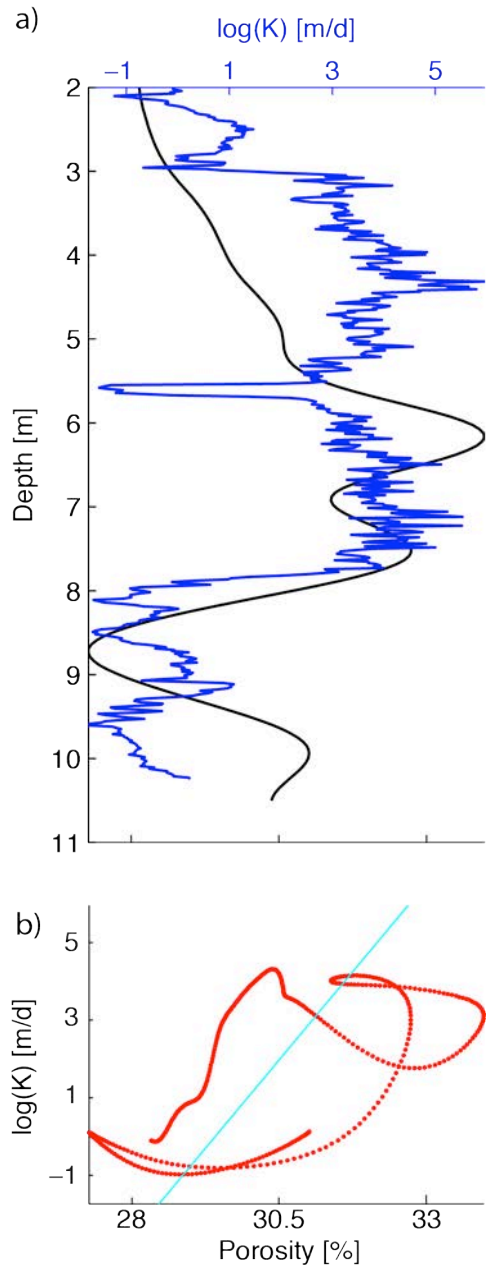

**Figure 6:** (a) High-resolution ln*K* (blue) and the porosity estimates from the original tomogram (black) at the transmitter borehole (borehole ID 091010B). (b) Scatter plot of smoothed lnK and porosity (red) together with the linear regression model in equation 4 (cyan).

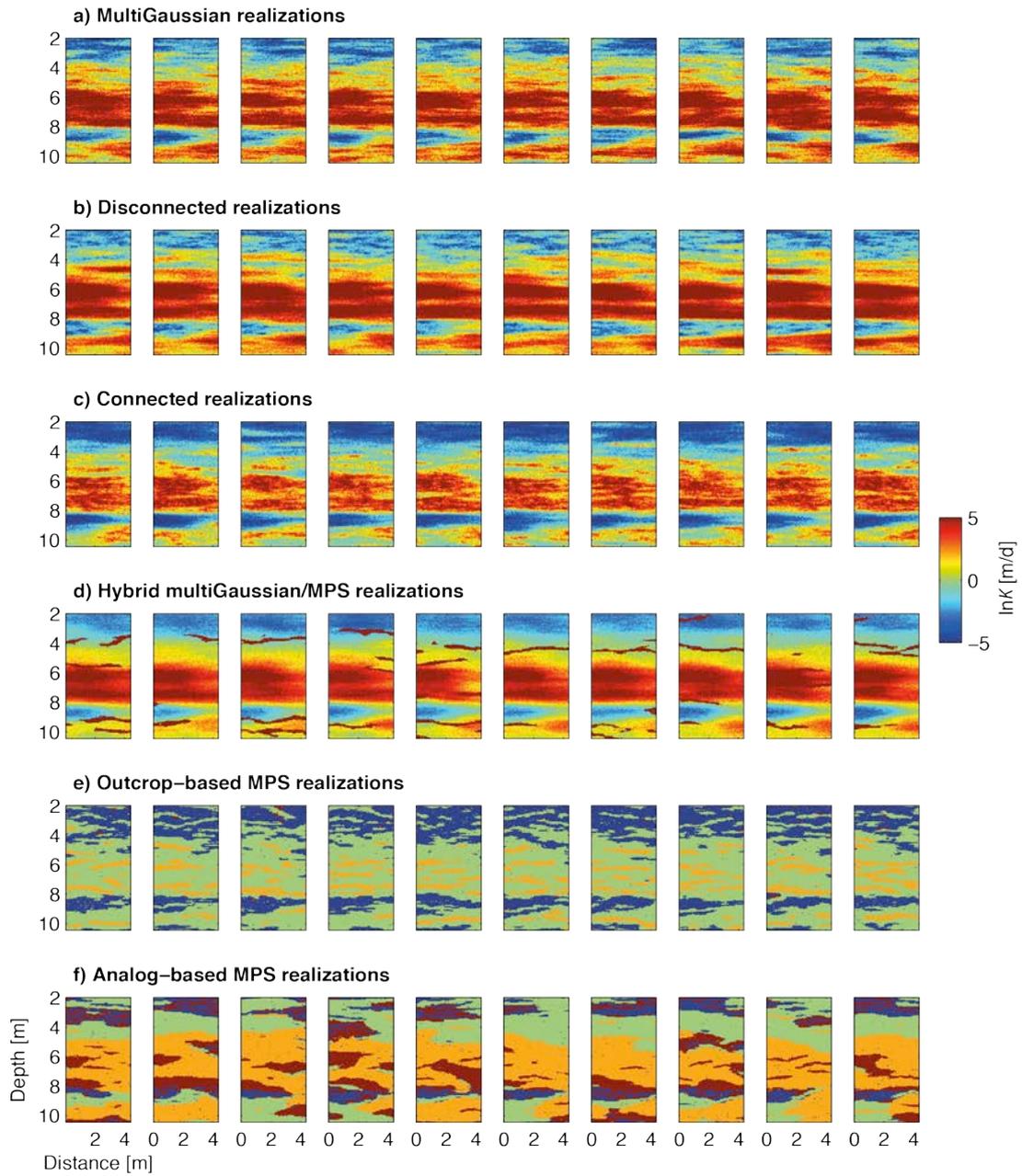

**Figure 7:** Realizations for the different geostatistical models, conditioned to the original crosshole GPR tomogram.

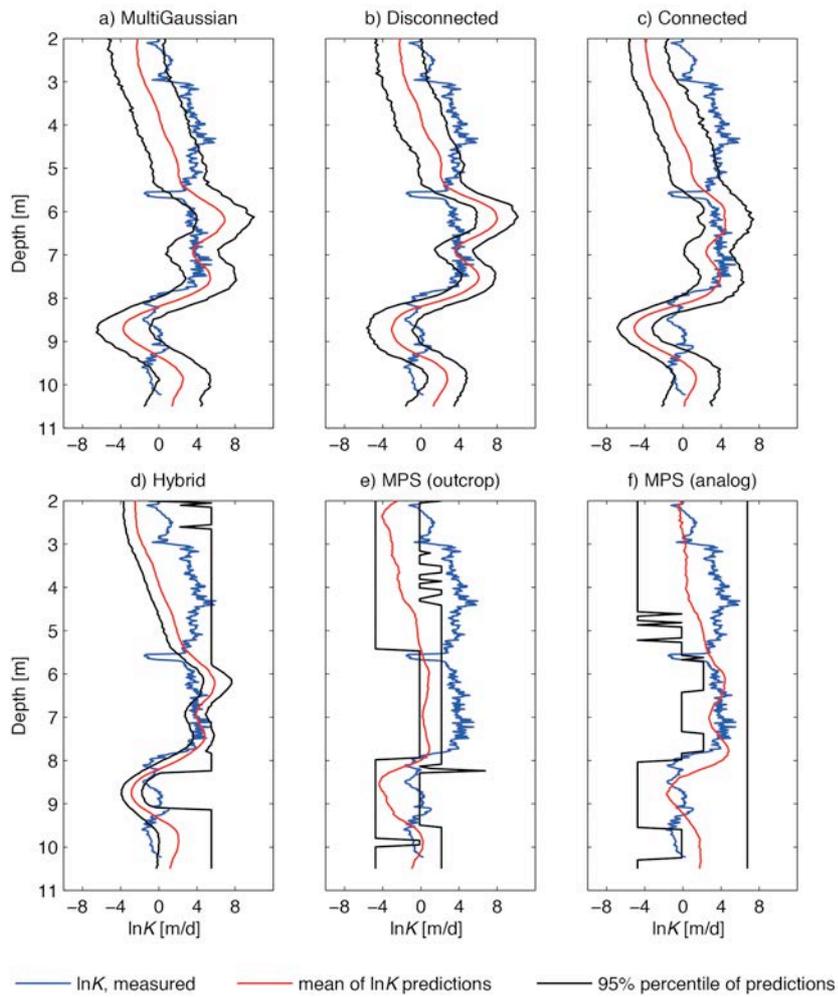

**Figure 8:** Measured ln*K* values (blue lines), the mean of predictions of ln*K* of the individual conditioned realizations for each geostatistical model (red lines) and the 95% percentile range of the predictions (black lines).

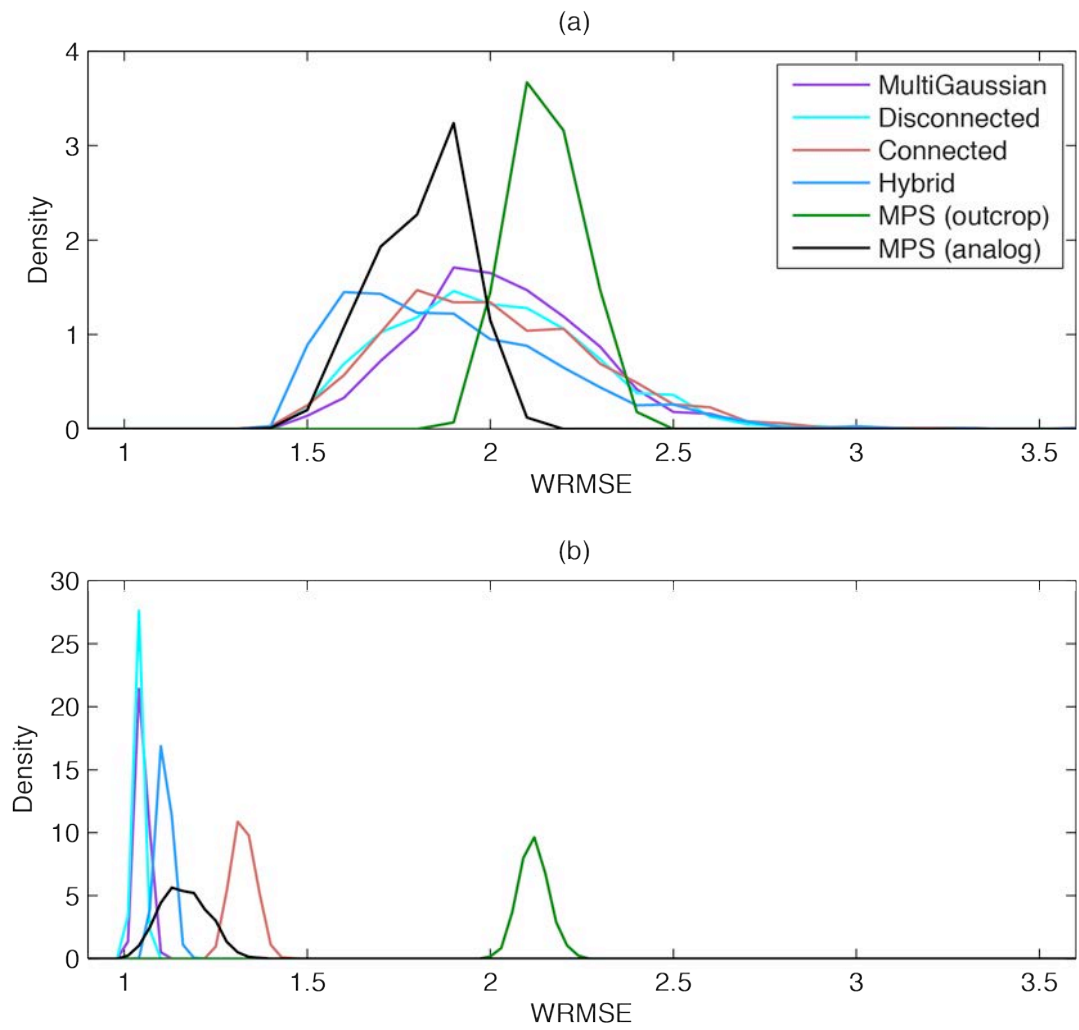

**Figure 9:** Probability density of the data fit for the (a) unconditional and (b) conditional realizations.

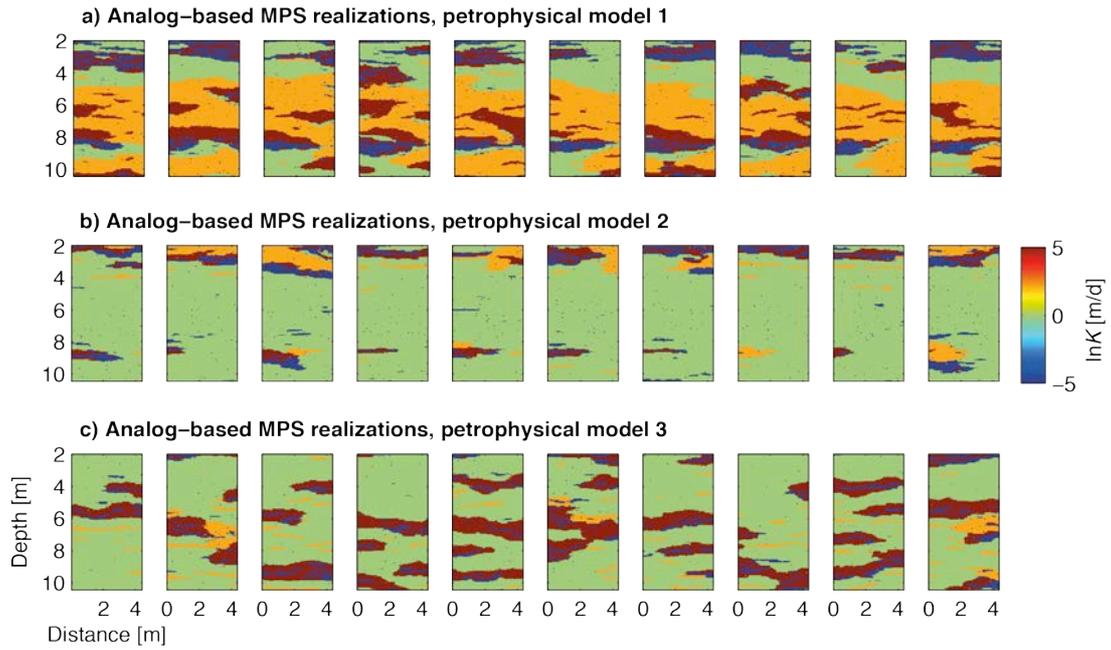

**Figure 10:** Realizations for the analog-based MPS model conditional to the original crosshole GPR tomogram for alternative petrophysical models. (a) The conditional realizations based on the site-specific petrophysical relationship (eq. 4) are the same as the ones shown in Figure 7f. (b) Conditional realizations for petrophysical model 2, in which a negative correlation between hydraulic conductivity and porosity was assumed. (c) Conditional realizations for petrophysical model 3, in which porosity and conductivity values were taken from Heinz et al. (2003).

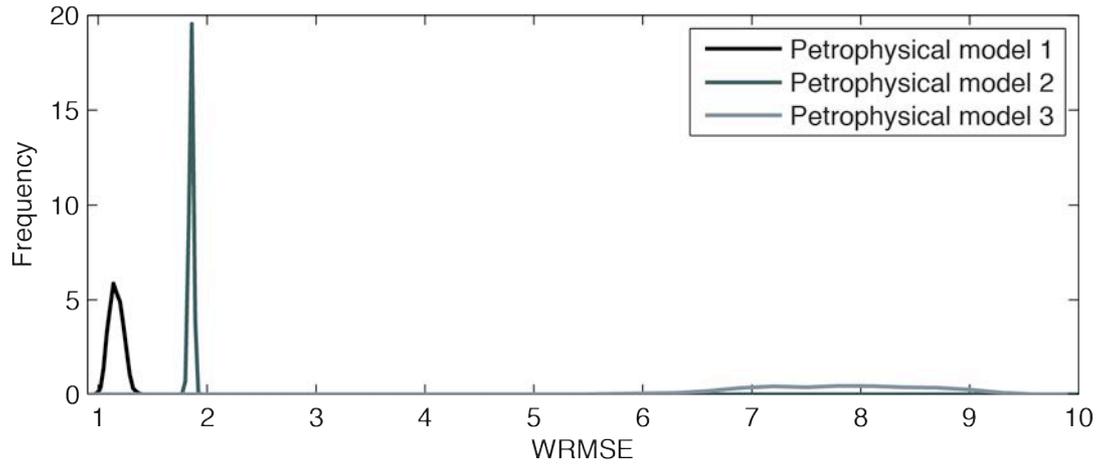

**Figure 11:** Probability density of the data fit for the conditional realizations of the analog-based MPS model for the three alternative petrophysical models.

**Table 1:** Overview of hydrogeological facies, their hydraulic conductivities (from Rehfeldt et al., 1992) and their derived porosities (from Heinz et al., 2003).

| Hydrofacies | ln$K$ [m/d] | $\varphi$ (%) |
|---|---|---|
| Undifferentiated sandy gravel | -0.15 | 20 |
| Sand | 2.16 | 43 |
| Sandy, clayey gravel | -4.75 | 17 |
| Open framework gravel | 6.76 | 31 |

**Table 2:** Overview of direct sampling parameters. The summation signs denote that the simple sum of non-matching nodes is taken as a distance measure. The scan fraction $f$ is a parameter for the multivariate TI and is not chosen for each individual variable.

| **Variable** | $n$ | $t$ | $f$ | **distance type** |
|---|---|---|---|---|
| Multi-Gaussian cases (continuous target variable) | | | | |
| tomogram | 25 | 0.04 | 0.1 | L1-norm |
| ln$K$ | 25 | 0.04 | 0.1 | L1-norm |
| MPS cases (categorical target variable) | | | | |
| tomogram | 25 | 0.04 | 0.1 | L1-norm |
| ln$K$ | 25 | 0.04 | 0.1 | $\sum$ |

**Table 3:** Numerical values of the maximum likelihood estimate for the various model types.

| Geostatistical model | Petrophysical Model | $\ln(\hat{L})$ ($\times 10^3$) |
|---|---|---|
| Multi-Gaussian | 1 | -1.72 |
| Disconnected | 1 | -1.72 |
| Connected | 1 | -1.95 |
| Hybrid multi-Gaussian/MPS | 1 | -1.76 |
| Outcrop-based MPS | 1 | -3.16 |
| Analog-based MPS | 1 | -1.71 |
| Analog-based MPS | 2 | -2.75 |
| Analog-based MPS | 3 | -16.37 |